\documentclass[runningheads]{llncs}
\usepackage{graphicx}
\usepackage{bbding}
\usepackage{gensymb}
\usepackage{algorithm}
\usepackage{algorithmic}
\usepackage{mathtools}
\usepackage{amsmath}
\usepackage{amssymb}
\usepackage{xcolor}
\usepackage{url}
\usepackage[normalem]{ulem}
\newcommand\redsout{\bgroup\markoverwith{\textcolor{red}{\rule[0.5ex]{2pt}{1pt}}}\ULon}

\begin{document}
\title{Bridge the Domain Gap Between Ultra-wide-field and Traditional Fundus Images via Adversarial Domain Adaptation}
%
%
\author{Lie Ju\inst{1,2}, Xin Wang\inst{1}, Quan Zhou\inst{3}, Hu Zhu\inst{3}, Mehrtash Harandi\inst{2}, \\ Paul Bonnington\inst{2}, Tom Drummond\inst{2}, and 
Zongyuan Ge\inst{1,2(}\Envelope\inst{)}}
\authorrunning{L. Ju et al.}
\titlerunning{Bridge the Domain Gap}
%
\institute{Airdoc LLC, China \and
Medical AI Group, Monash University, Australia \\
\url{(https://mmai.group)}
\\
\and Nanjing University of Posts and Telecommunications, China
\\ 
\email{julie@airdoc.com}, \email{zongyuan.ge@monash.edu}}
\maketitle              
\begin{abstract}
For decades, advances in retinal imaging technology have enabled effective diagnosis and management of retinal disease using fundus cameras. Recently, ultra-wide-field (UWF) fundus imaging by Optos camera is gradually put into use because of its broader insights on fundus for some lesions that are not typically seen in traditional fundus images. Research on traditional fundus images is an active topic but studies on UWF fundus images are few. One of the most important reasons is that UWF fundus images are hard to obtain. In this paper, for the first time, we explore domain adaptation from the traditional fundus to UWF fundus images. We propose a flexible framework to bridge the domain gap between two domains and co-train a UWF fundus diagnosis model by pseudo-labelling and adversarial learning. We design a regularisation technique to regulate the domain adaptation. Also, we apply MixUp to overcome the over-fitting issue from incorrect generated pseudo-labels. Our experimental results on either single or both domains demonstrate that the proposed method can well adapt and transfer the knowledge from traditional fundus images to UWF fundus images and improve the performance of retinal disease recognition.

\end{abstract}
\section{Introduction}
\begin{figure}
	\includegraphics[width=12cm]{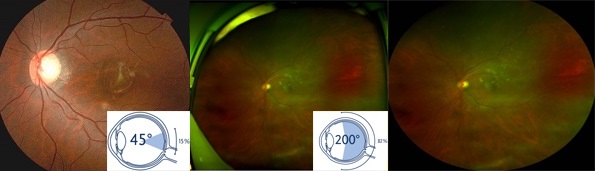}
	\centering
	\caption{Left: traditional fundus image. Middle: Optos UWF fundus image. Right: Optos UWF fundus image after preprocessing.} \label{fig1}
\end{figure}

Fundus screening can detect abnormal retinal diseases such as diabetic retinopathy (DR), age-related macular degeneration (AMD) and glaucoma. It also provides a relatively good prognosis of visual acuity at the early stage. However, cataract, vitreous opacity and other diseases with weak refractive stroma are often difficult to be imaged via traditional examinations because of obstruction on the optical path. In the 2000s, Optos ultra-wide-field (UWF) fundus imaging first became commercially available, the image's capture range can cover 80\% - 200\degree of the retina, compared to only 30\degree - 60\degree achieved with traditional retinal cameras. As Fig.~\ref{fig1} shows, UWF imaging covers a greater retinal area, allowing more clinically relevant pathology, which usually changes from the peripheral retina to be detected, such as retinal degeneration, detachment, haemorrhages, exudations and so on~\cite{kiss2014ultra}.


For years, deep learning has been widely used on traditional fundus images and achieved good performance on diagnosis various retinal diseases~\cite{fu2018disc,wang2019two,wang2019retinal}, but few cases are being seen on UWF fundus images~\cite{nagasato2018deep,ohsugi2017accuracy}. In \cite{nagasato2018deep}, they used deep learning (DL) and
support vector machine (SVM) algorithms to detect central retinal vein occlusion (CRVO) in UWF images. And \cite{ohsugi2017accuracy} applied those methods on detecting rhegmatogenous retinal detachment (RRD). However, 
we find two main limitations of the existing UWF related works. Firstly, the dataset being trained and tested only contain clean and ideal samples collected from a controlled environment. In real scenarios, the UWF fundus images are often interfered by the eyelids and eyelashes. Those interferents may affect the screening performance of the model trained on clean images. Secondly, since the UWF fundus datasets are scarce and difficult to obtain, most existing literature studies one specific disease only, and the universality of the algorithm to various retinal diseases is not always guaranteed. Therefore, it is important to have a technique which can not only leverage the available traditional fundus images for co-training but also transfer the knowledge from the well-performed traditional retinal-based disease recognition model to UWF model.  



In this work, we propose a novel multi-disease diagnosis framework for UWF, which can take advantage of the existing public fundus images, and transfer abstract knowledge from one domain to another. 
To achieve that, we employ generative adversarial network (GAN) to map fundus images into UWF images. In addition, we use pseudo-label~\cite{lee2013pseudo} technique to generate labels for transferred fundus images, and employ MixUp~\cite{zhang2017mixup} to calibrate the incorrect predictions by pseudo-labelling.
To the best of our knowledge, we are the first work to study the probability-based knowledge transfer between the UWF and traditional fundus images under Deep Convolutional Neural Network (DCNN) framework.
We evaluate our proposed method of UWF images from the real clinical setting. The experimental results demonstrate that our proposed method indeed contributes to classification performance improvement on common lesions such as haemorrhage, drusen etc., and well bridge the domain gap between UWF and traditional fundus images for model co-training.





\section{Datasets}
\subsubsection{Data Annotation}

Our two domains of the dataset (traditional as \textit{source} and UWF as \textit{target}) were acquired from private hospitals and each image was labelled by three ophthalmologists. The image will preserve only if at least two ophthalmologists are in agreement of the disease labels. For better universality and generalisation, we select five kinds of lesions which are most common in two different types of fundus images: 1) Hemorrhage, 2) Drusen, 3) Hard Exudation, 4) Soft Exudation, 5) Retinal Hole. 
The details of each lesion category are shown in Table.~\ref{Table1}. 
The dataset is randomly divided into training (50\%), validating (25\%) and testing (25\%) images.
\begin{table}[]
	\centering
	\setlength{\tabcolsep}{3mm}
	\caption{Data statistics. About 32\% of images have dual-class or more labels.}
	\begin{tabular}{c|c|c|c|c|c|c}
		\hline
		       & Normal  & Hemorrhage & Drusen & HE  &  SE   & RH   \\ \hline
		Fundus (\textit{Source})  & 800 & 3047 & 1472 & 679 & 560 & 60  \\
		UWF (\textit{Target}) & 122 & 648  & 198  & 393 & 143 & 134 \\ \hline
	\end{tabular}
	\label{Table1}
\end{table}

\subsubsection{Data Preprocessing}
To remove the interferents such as eyelids and eyelashes from the UWF images, we use Otsu segmentation~\cite{otsu1979threshold} to first locate the obstructions at the border of the image. Then we use zero value to mask out these pixels, only to keep the elliptical part from the fundus as shown in the right part of Fig.~\ref{fig1}. For some more difficult cases, we train a U-Net~\cite{ronneberger2015u} segmentation method with manual annotations. See our supplementary material for more information.



\section{Methods}
\subsubsection{Problem Definition}Let $X^{S} = \{x_{1}^{S},x_{2}^{S},...,x_{N_{S}}^{S}\}$ and $X^{T} = \{x_{1}^{T},x_{2}^{T},...,x_{N_{T}}^{T}\}$ denote the source domain (traditional) and target domain (UWF) images respectively. The corresponding labels are defined as $Y^{S} = \{y_{1}^{S},y_{2}^{S},...,y_{N_{S}}^{S}\}$ and $Y^{T} = \{y_{1}^{T},y_{2}^{T},...,y_{N_{T}}^{T}\}$. 
Our goal is to map the fundus images with abundant annotations $X^{S}$ into target domain UWF images. The generated UWF images are called as pseudo target samples $\hat{X}^{T}$, along with pseudo-labels $\hat{Y}^{T}$ are used to assist the model co-training in UWF domain.


\subsubsection{Method Overview}
\begin{figure}[!t]
	\includegraphics[width=12cm]{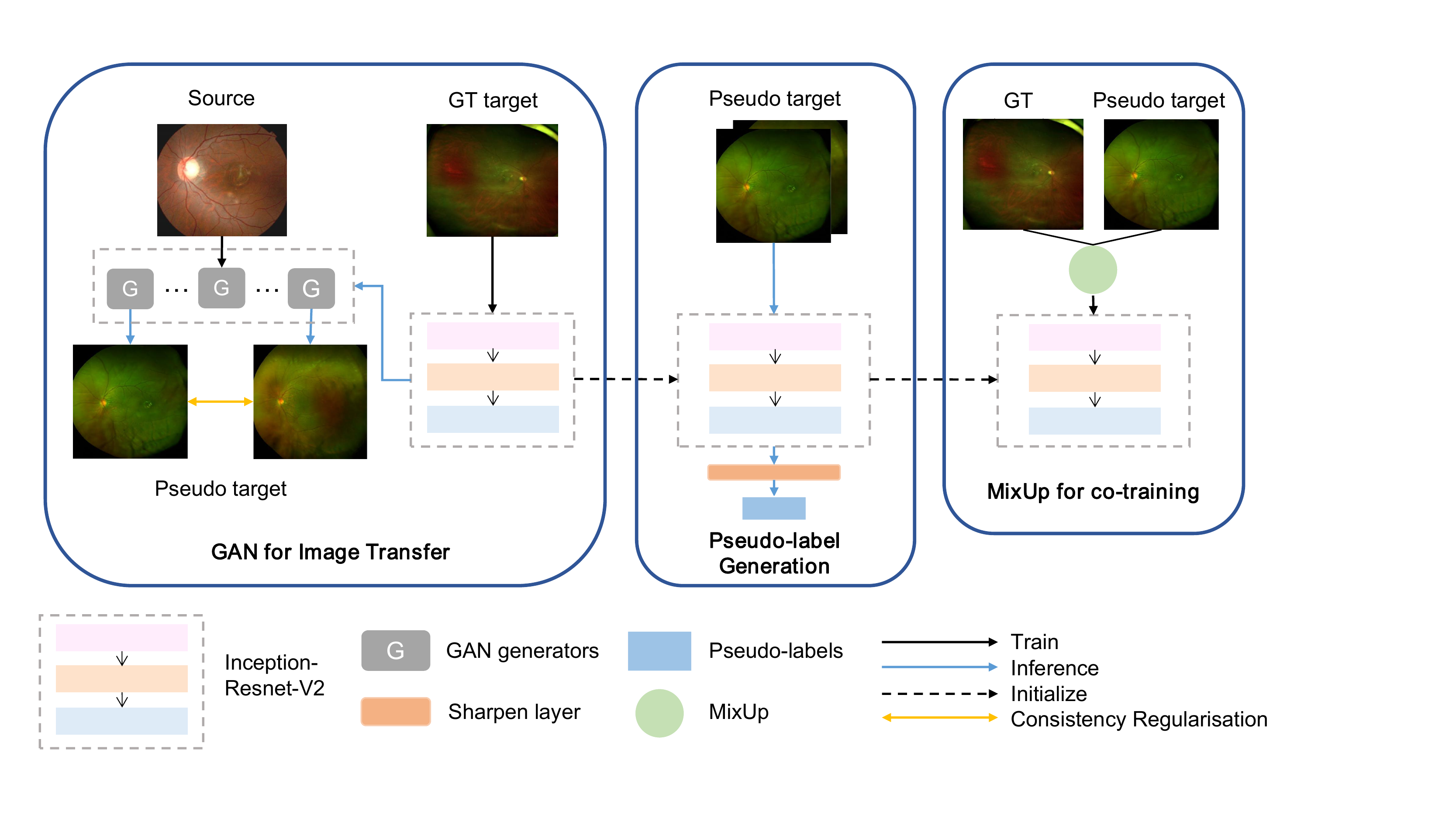}
	\centering
	\caption{Overview of the proposed method.} \label{fig2}
\end{figure}

The overview of our proposed method is shown in Fig.~\ref{fig2} and can be divided into three stages. 
During the first stage, we train a GAN model to transfer fundus images into UWF fundus images. We also train a target domain classification network to assist the GAN training as a discriminator and produce pseudo-labels at the later stage. 
At the second stage, we generate target domain pseudo-labels $\hat{Y}^{T}$ for transferred images $\hat{X}^{T}$. 
Finally, we use the GT target images along with the pseudo target images (with pseudo-labels) to co-train the domain-adapted target image classifier.
In the following sections, we will describe our GAN mapping model for source-to-target image mapping, pesudo-label generation and final stage co-training using MixUp in details. 


\subsection{GAN for Image Transfer}
\label{stage1}
To utilise samples from the source domain and co-train them at the target domain, we need to first map data from one domain to the other. 
We employ Cycle-GAN~\cite{zhu2017unpaired} as our generator. Cycle-GAN has two generator-discriminator pairs, and it can ensure that the essential class features can be preserved using the Cycle-GAN loss. Moreover, Cycle-GAN has no requirement on paired images with labels. 
Our motivation of using the source domain colour fundus images to assist target domain UWF images training is based on the factor that although there are great differences in colour and style between those two domains, the clinical definition and diagnostic criteria of lesions are the same~\cite{nagiel2016ultra}. Therefore, there exists a shared, cross-domain feature space to represent the same disease across two domains. And the extracted features and knowledge from the source domain can be learned by the target domain task after adaptation.


\vspace{-1em}


\subsubsection{Consistency Regularisation }
To further preserve pathology features and regulate the quality of pseudo target images during the domain image transformation, we aim to train a series of generators with consistency regularisation applied for these generators.
Consistency regularisation is gradually regarded as a gold-standard technique in semi-supervised learning~\cite{hoang2020semixup,laine2016temporal,sajjadi2016regularization,xie2019unsupervised}. The idea is heuristic: under the condition of not destroying the semantic information, the input images for training are randomly flipped, cropped or transformed by data augmentation operations, and an additional penalty is added to the loss function. 
\begin{figure}[!t]
	\includegraphics[width=12cm]{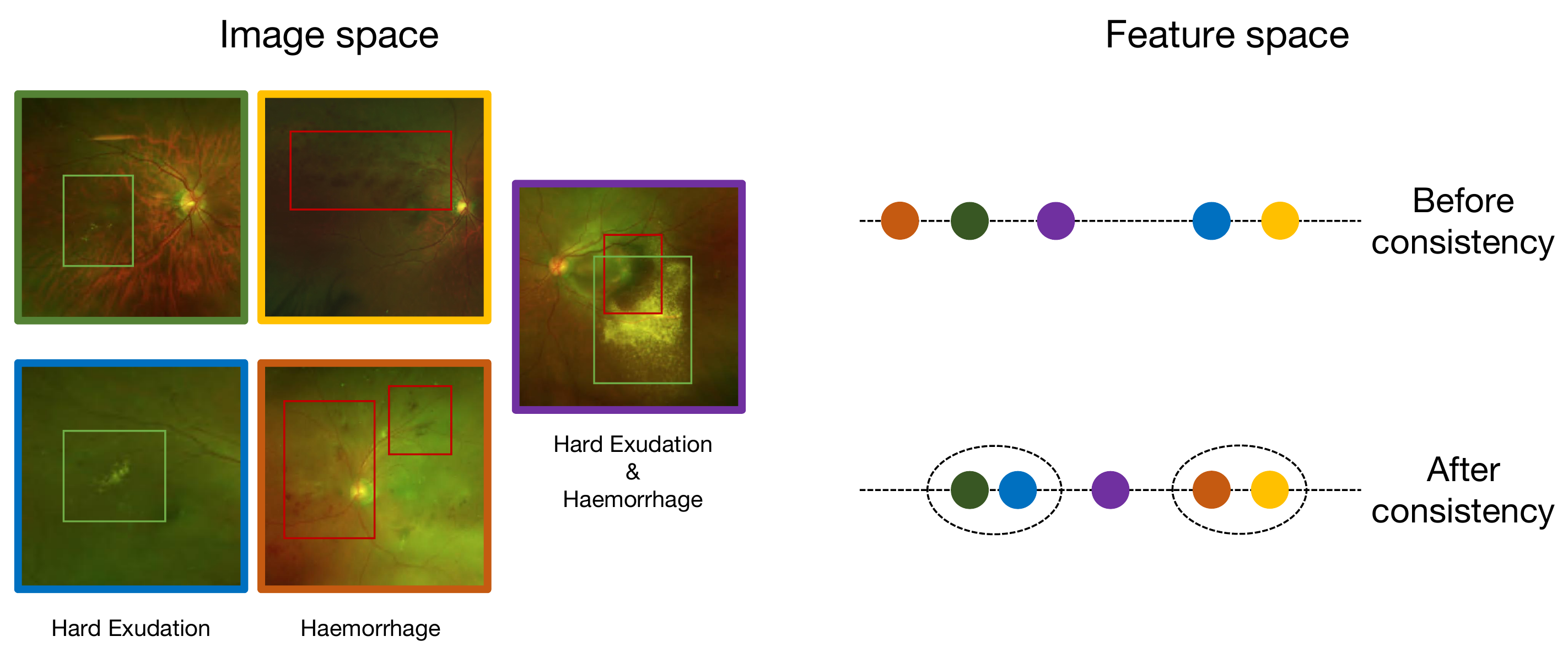}
	\centering
	\caption{An illustration of consistency regularisation. On the left of the figure, hard exudation and haemorrhage are marked with green and red bounding boxes respectively. The outline colour of the images corresponds to the colour dots on the right. We can observe that after applying consistency regularisation, images from the same category are being pulled closer in feature space. Images containing both categories also benefit from this trend.} \label{fig3}
\end{figure}
\cite{zhang2019consistency} proposed to add consistency regularisation for GANs, which aims to enforce the discriminator to be unchanged by arbitrary semantic-preserving perturbations and to focus more on semantic and structural changes between real and fake data. In our work, it well regulates the images in the feature space as Fig.~\ref{fig3} shows.
We defined our consistency regularisation term as follow:
\begin{equation}
\begin{split}
    r &= Randint([1, k)) \, \,  k \in [2, K], \\
    L_{cr} &= \left \| h_{T}(G_{S \rightarrow T,r}(x^{S}|\theta_{r})|\theta_{T}) - h_{T}(G_{S \rightarrow T,k}(Aug(x^{S})|\theta_{k})|\theta_{T}) \right \|^{2},
\end{split}
\end{equation}
where $G_{S \rightarrow T}$ is the generator that transfers the source images into target images. $Aug(x)$ denotes a stochastic data augmentation function. $K$ is a hyperparameter of the number of generators being trained. $G_{S \rightarrow T,r}(x^{S})$ means we randomly pick a generator from the trained generator pool. $\theta_{r}$ and $\theta_{k}$ denote the parameters of a randomly picked generator and the current training generator, respectively. For each sample, we input $x^{S}$ and $\hat{x}^{S}$ generated by $G_{S \rightarrow T,r}(x^{S}|\theta_{r})$ and $G_{S \rightarrow T,k}(\hat{x}^{S}|\theta_{k})$ into the classification network $h_{T}(x|\theta{T})$ trained by $X^{T}$ with GT label. 
The total loss in adversarial training can be expressed as:
\begin{equation}
\begin{split}
L_{total} &= L_{Cycle-GAN} + \lambda_{cr} * L_{cr},
\end{split}
\end{equation}
where hyper-parameters $\lambda_{cr}$ and $K$ is set as $\lambda_{cr}$ = 1 and $K$ = 2 in our experiments. For more details about $L_{Cycle-GAN}$,  please refer to~\cite{zhu2017unpaired}. 

\subsection{Pseudo-label Generation}
\label{sec. pseudo-label}
Using pseudo-labelling to "guess" towards true labels for unlabelled or generated data is widely used in semi-supervised learning~\cite{berthelot2019mixmatch,ding2019feature,usunier2011multiview,xing2019adversarial}. In semi-supervised learning, a common assumption is that the decision boundary of classifier should not pass through high-density regions of the marginal data distribution, and the distribution gap while learning discriminative features for the task can be reduced~\cite{berthelot2019mixmatch,saito2019semi}.
To achieve this, we propose to add a sharpen layer after the Sigmoid layer for unlabelled data (see pseudo-label generation in Fig.~\ref{fig2}) to produce the predictions with minimised entropy. 
We use the cumulative distribution function of the normal distribution as our sharpen function, which is defined as:
\begin{equation}
\label{equation_sharpen}
\mathrm{Sharpen}(y_{i}) = \frac{1}{2}(1+\textrm{erf}(\frac{y_{i}-\mu }{\sigma \sqrt{2}})),
\end{equation}
 where $y_{i}$ is $i$th class in one-hot encodings, and erf denotes the Gaussian error function. We set $\mu=0.4$ and $\sigma^{2}=0.01$ in this work.


\subsection{MixUp for co-training}
\label{sec. mixup}

In the last stage, we mix pseudo and GT target images $\{{X}^{T},\hat{X}^{T}\}$ and then shuffle those samples for co-training. However, the wrong information of pseudo-labels will mislead the training, which is known as confirmation bias~\cite{arazo2019pseudo}. Recently, MixUp~\cite{zhang2017mixup} is regarded as a strong regularisation technique and calibrate the network to favour linear behaviour in-between training samples, reducing oscillations in regions far from them.
The process of MixUp can be defined as:
\begin{equation}
\begin{split}
\tilde{x} &= \lambda*x_{i} + (1-\lambda)*x_{j}, \\
\tilde{y} &= \lambda*y_{i} + (1-\lambda)*y_{j},
\end{split}
\end{equation}
where $x_{i}, x_{j}$ are raw input vectors, and $y_{i}, y_{j}$ are corresponding one-hot label encodings.  In this work, the hyperparameter $\lambda$ is set as 0.4.

\section{Experiments}
\subsubsection{Implementation Details}All training and testing images are resized to $499 \times 499$. For the classification network training, we use the “Inception-ResNetV2” model from~\cite{szegedy2017inception} as the backbone model with pre-trained parameters.
We apply ADAM~\cite{kingma2014adam} to optimize all models. The learning rate starts at 1e-3 and changed to 3e-4 after 10 epochs. For evaluation metrics, we employ accuracy, precision, recall, specificity and F-measure. Results are reported on 4-fold cross-validation.


\begin{figure}[]
	\includegraphics[width=12cm]{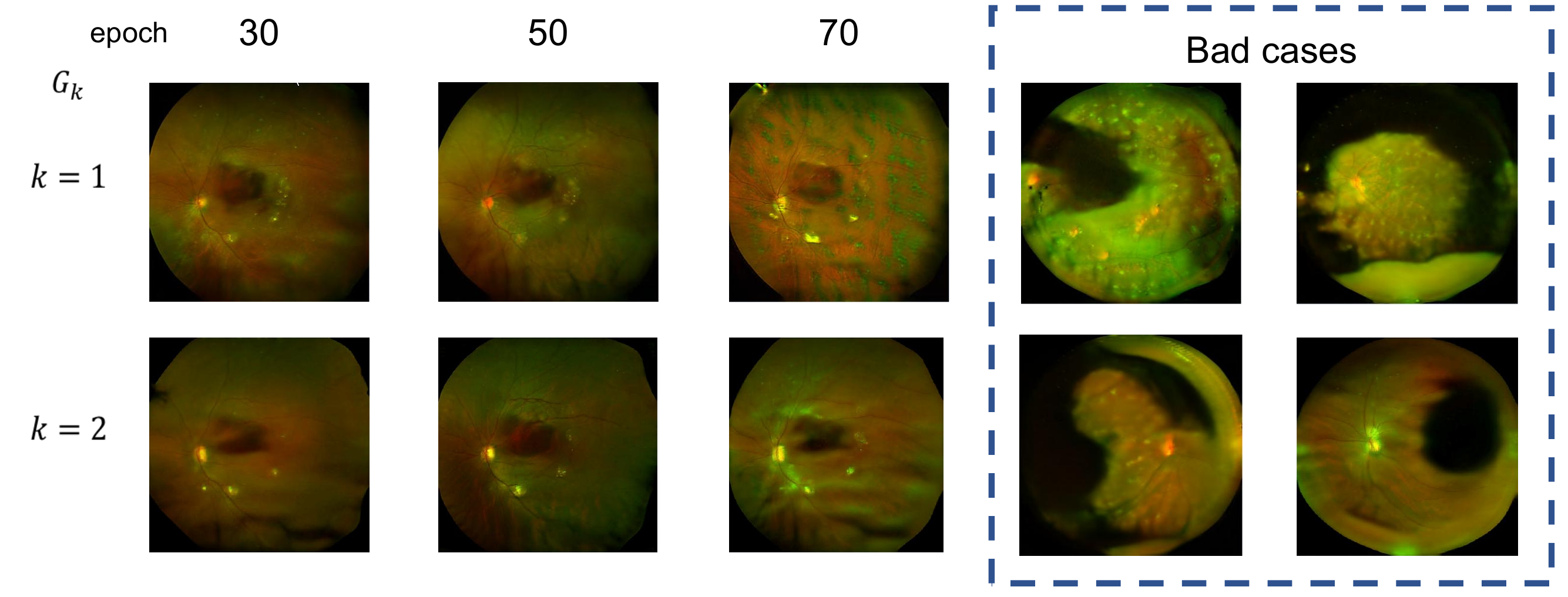}
	\centering
	\caption{Illustrations of generated pseudo target images from different epochs and generators.}
	\label{fig4}
\end{figure}

\subsubsection{Pseudo Target Images Analysis}Fig.~\ref{fig4} shows some examples of generated UWF (pseudo target images) transferred from fundus images. It can be seen that there exists a coarse to fine refinement process when more epochs are being trained.  
However, at the 70th epoch, we observe interferents and noise all over the fundus region. This is due to the overfitting to the noise of GT target images.
We also compare the output of two different generators (row 1 vs row 2). Although the form of interferents is uncontrollable (green spots in $k = 1$, cotton-like objects in $k = 2$), the lesions and pathology features are consistent. It indicates that the proposed consistency regularisation enforcing the generated images to have close predictions and semantic information we care about.  
Some bad cases caused by laser spots after surgery, eyelids not completely removed etc. are shown in the right-most two columns. 
\begin{table}[!t]
\centering
\setlength{\tabcolsep}{2.5mm}
\caption{Overall results of our proposed methods.}
\begin{tabular}{l|c|c|c|c|c}
\hline
Methods    & Accuracy         & Precision        & Recall           & Specificity      & F1-score        \\ \hline
UWF              & 75.55\%          & 58.95\%          & 52.89\%          & \textbf{84.80\%} & 55.47\%          \\
Fine-tuning   & \textbf{77.02}\%          & \textbf{59.91}\%          & \textbf{61.83}\%          & 83.19\%          & \textbf{60.84}\% \\ \hline
KD~\cite{hinton2015distilling}         & 74.30\%          & 54.89\%          & 58.09\%          & 80.81\%          & 56.44\%          \\
Pseudo-label~\cite{lee2013pseudo}        & 77.74\%          & 61.85\%          & 60.22\%          & \textbf{84.84}\%          & 60.92\%          \\
MixUp~\cite{zhang2017mixup}                & 77.03\%          & 59.21\%          & 63.93\%          & 82.30\%          & 61.48\%          \\

Ours (3309 $\times$ K)        & \textbf{78.73\%} & \textbf{62.94\%} & \textbf{64.70\%} & 84.40\%          & \textbf{63.75\%} \\ \hline
Ours (2500 $\times$ K)       & 79.75\%          & 65.35\%          & 62.53\%          & 86.67\%          & 63.91\%          \\ 
Ours (1500 $\times$ K)       & 78.27\%          & 60.97\%          & \textbf{67.57}\%          & 82.57\%          & 64.07\%          \\
Ours (500 $\times$ K)        & \textbf{81.37\%} & \textbf{68.39\%} & 65.12\%          & \textbf{87.90\%} & \textbf{66.71\%} \\

\hline
\end{tabular}
\label{Table2}
\end{table}

\subsubsection{Baseline Models}
In this study, we consider two baselines: 1) \textbf{UWF:} Training from GT target images. 
2) \textbf{Fine-tuning:} Training from GT source fundus images and then fine-tuning on the GT target UWF images. The results are shown in the top section of Table.~\ref{Table2}. The results of \textbf{Fine-tuning} demonstrates improvements in accuracy, precision, specificity and F1 score by 1.47\%, 0.96\%, 8.94\%, and 5.37\% respectively compared with trained using UWF fundus images only, which suggests that there exists an effective shared, cross-domain feature space to represent the same disease across two imaging domains.

\subsubsection{Comparative and Ablation Studies}To better quantify the performance of knowledge transfer between two domains and have a comparison study of our proposed method, we implement three popular knowledge-based learning models. 
\textbf{KD:} The knowledge distillation (\textbf{KD})~\cite{hinton2015distilling} technique\footnote{In our scenario, we set the model trained from source images as our teacher model and that trained from target images as our student model.}.  
\textbf{Pseudo-label:} We transfer annotation knowledge for pseudo target images (K=1, without consistency regularisation) through pseudo-labelling, and train the classification using pseudo target images and GT target images. 
\textbf{MixUp:} Beyond the pseudo-label training above, we further apply MixUp for training. 

The middle part of Table.~\ref{Table2} shows the results of our comparative study. 
Knowledge distillation gets worse performance than the \textbf{Fine-tuning} baseline. It indicates that although teacher and student models share similar classification tasks and features, the domain knowledge gap between UWF and fundus images are not easy to adapt and that the knowledge learned from teacher model is not very effective for training a classifier.
Both pseudo-label and MixUp gain marginal improvement compared with the baseline in term of F1-score by 0.08\% and 0.64\% respectively. 
It is reassuring to see our proposed method achieves the best performance on this problem. Almost comprehensive improvements are observed, especially compared with the baseline trained from UWF fundus images, on the accuracy, precision, recall and F1-score by 3.18\%, 3.99\%, 11.81\%, and 8.28\% respectively. 


In the last part of the table, we conduct an ablation study by reporting the corresponding performance of the number of added pseudo target images during co-training. It is surprising to observe that more pseudo target images do not guarantee better performance. It suggests that the quality of generated pseudo target images matters but is difficult to regulate. Too many low-quality images (right part of Fig.~\ref{fig4}) will instead harm the training of classification network. 


\section{Conclusion}
In this paper, for the first time, we studied the domain characteristics between UWF and traditional fundus images and proposed a novel framework to bridge this gap. It adapts traditional fundus images to UWF fundus images using cycle-consistent adversarial learning with an effective consistency regularisation term. We also applied MixUp to minimise the confirmation bias and avoid over-fitting for the network co-training. Our experiments demonstrated that our proposed method can significantly improve the performance of recognition lesions on UWF fundus images, which has opened more possibilities for the research on automatic screening of UWF fundus images.
\clearpage
\bibliographystyle{splncs04.bst}
\bibliography{samplepaper.bib}
\clearpage

\section{Appendix}

\subsection{Preprocessing Segmentation Model Training: }
\begin{figure}[!]
	\includegraphics[width=11cm]{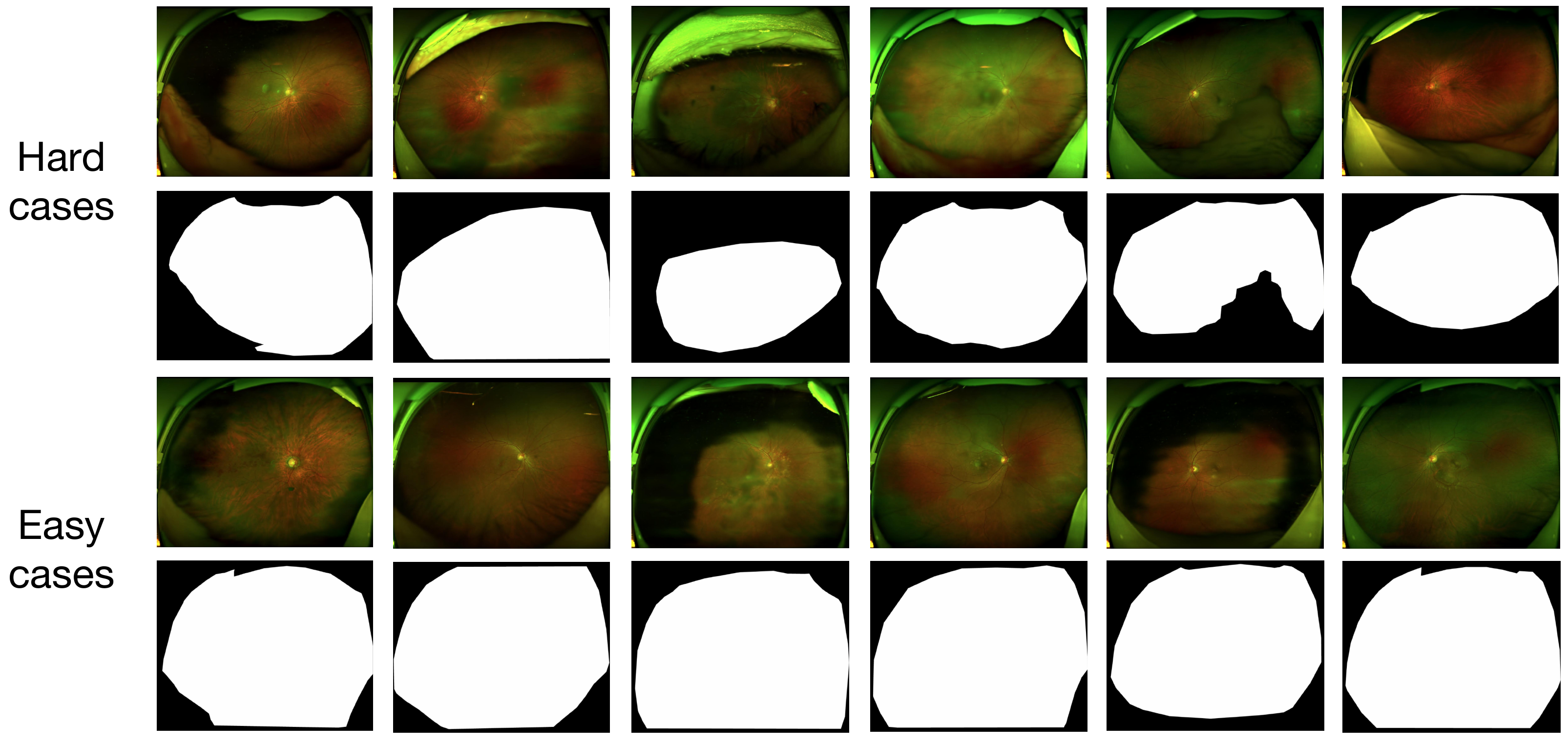}
	\centering
	\caption{Annotations for interferents.}
	\label{fig5}
\end{figure}

As Fig.~\ref{fig5} shows, for some easy cases, because the interferents are similar in colour and occupy a small proportion in the image, using simple unsupervised segmentation methods can remove them. However, for some bad cases, manual annotations and using deep learning method to achieve automatic segmentation is very necessary.

\begin{figure}[!]
	\includegraphics[width=8cm]{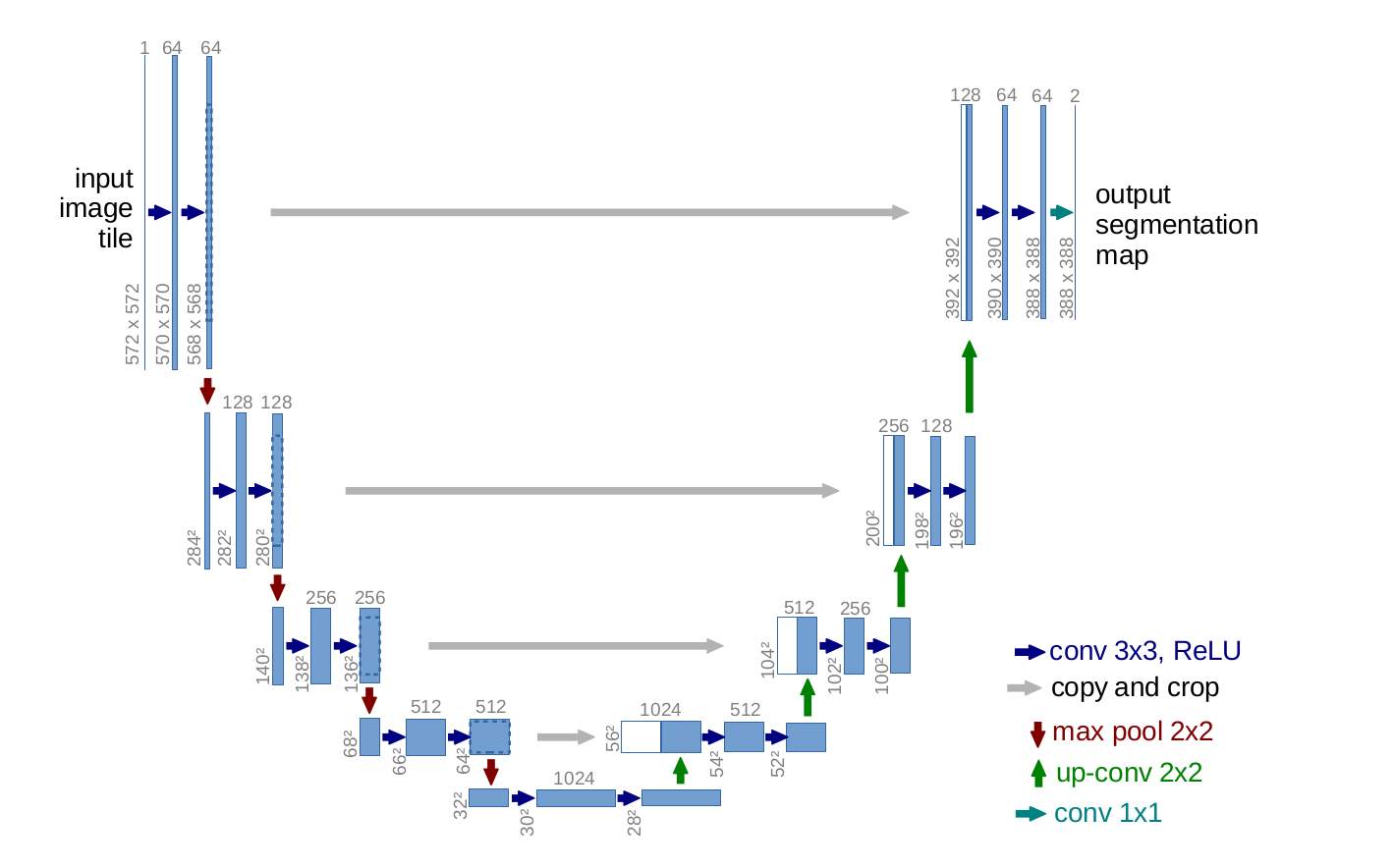}
	\centering
	\caption{The overview of U-net network~\cite{ronneberger2015u}.}
	\label{fig6}
\end{figure}

Compared with labelling lesions, labelling interferents does not require much expert knowledge. In this work, we selected 300 UWF fundus images from the datasets for annotation. We apply U-net network for segmentation network training and the structure of U-net~\cite{ronneberger2015u} is shown as Fig.~\ref{fig6}.

\vspace{-1cm}
\begin{figure}[!]
	\includegraphics[width=11cm]{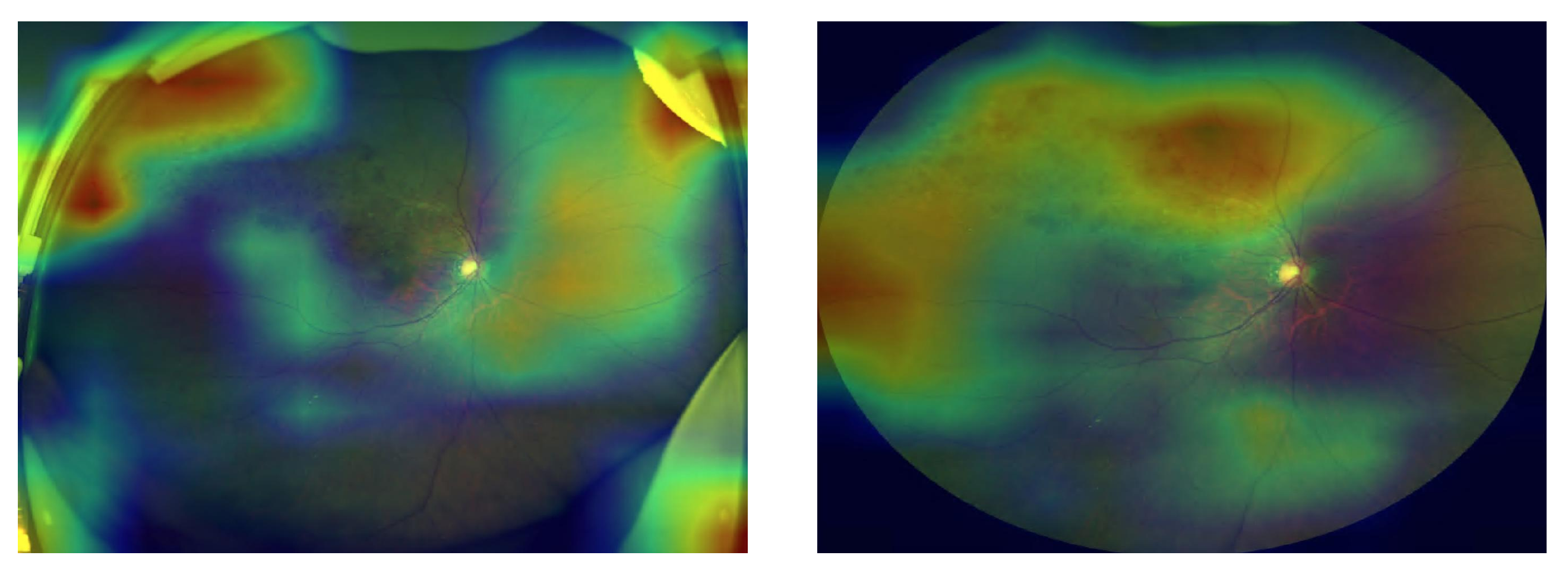}
	\centering
	\caption{The results of Grad-CAM.}
	\label{fig7}
\end{figure}

Gradient-weighted Class Activation Mapping (CAM)~\cite{selvaraju2017grad} technique allows the classification-trained CNN to localize class-specific image regions. Given two classification networks trained by images without and with preprocessing respectively. The image in Fig.~\ref{fig7} are predicted to be haemorrhage in both models. However, as the left part of the figure shows, the decision of the former model is affected by the interferents seriously, even though the prediction is correct. After preprocessing, the heat map is more focused on lesion areas, as the right part shows.

\subsection{Visualisation of Sharpen Function:}

\begin{figure}[!]
	\includegraphics[width=9cm]{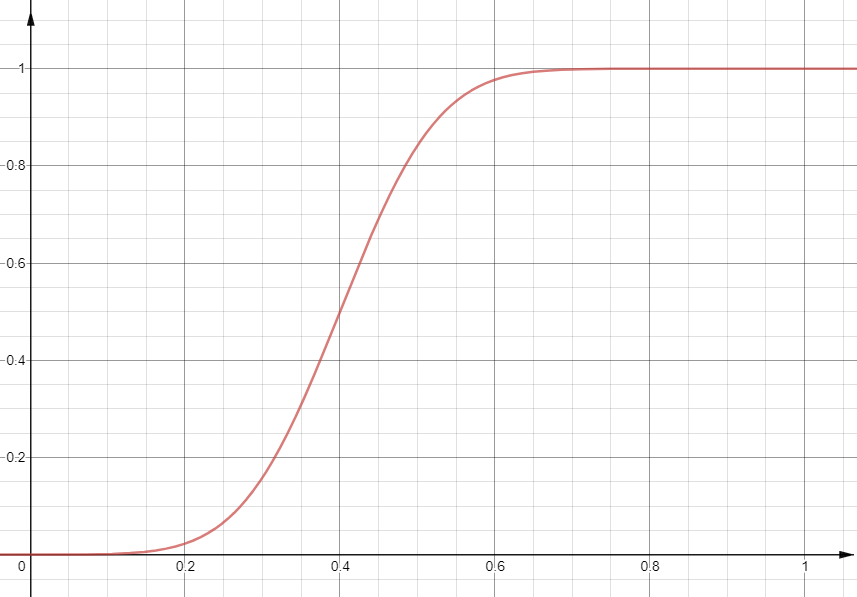}
	\centering
	\caption{Sharpen function Eq. \ref{equation_sharpen} when $\mu=0.4$ and $\sigma^{2}=0.01$.}
	\label{fig8}
\end{figure}

\clearpage
\subsection{Algorithm Table}
Algorithm. \ref{algorithm_1} gives details of our proposed methods. It should be noted that MixUp is regarded as an augmentation technique, so we need both GT target $X^{T}$ images and shuffled images after MixUp $X^{B}$ for the target classification network training.

\begin{algorithm}[]

	\caption{Bridge the domain gap}
	\label{algorithm_1}
	\begin{algorithmic}
		\REQUIRE ~~\\
		All the images $X^{S} = \{x_{1}^{S}, x_{2}^{S}, ..., x_{N_{S}}^{S}\}$, $X^{T} = \{ x_{1}^{T}, x_{2}^{T}, ..., x_{N_{T}}^{T}\}$ and corresponding labels $Y^{S} = \{y_{1}^{S}, y_{2}^{S}, ..., y_{N_{S}}^{S}\}$, 			$Y^{T} = \{y_{1}^{T}, y_{2}^{T}, ..., y_{N_{T}}^{T}\}$.

		\ENSURE ~~\\
		\textbf{Stage. 1 \& 2: }Generate pseudo target images and labels:
		
		Train classifier $h_{S}(X^{S} | \theta_{S})$ with $X^{S}$ and $Y^{S}$. 
		
		Train classifier $h_{T}(X^{T} | \theta_{T})$ with $X^{T}$ and $Y^{T}$. Where $\theta_{T}$ is initialized from $\theta_{S}$.
		
		Train generator $G_{1}(X^{S} | \theta_{1})$ with $X^{S}$ and $X^{T}$.
		
		Generate pseudo target labels $\hat{Y}_{1}^{T}$ = $h_{T}(\hat{X}_{1}^{T} | \theta_{T})$.

		\FOR{$k = 2$ \TO $K$ } 
		\STATE{
			$r = Random([1,k))$
			
			$\hat{X}^{S}$ = Aug($X^{S}$)
			
			Train $G_{k}(\hat{X}^{S} | \theta{k}, h_{T}(\hat{X}_{r}^{T} | \theta_{T}))$ with $\hat{X}^{S}$, $X^{T}$ and $h_{T}(X^{T} | \theta_{T})$.
		
			Generate pseudo target samples: $\hat{X}_{k}^{T}$ = $G_{k}(X^{S} | \theta{k})$.
			
			Generate pseudo target labels: $\hat{Y}_{k}^{T}$ = $h_{T}(\hat{X}_{k}^{T} | \theta_{T})$.

		} 
		
		\ENDFOR

		\textbf{Stage. 3:} MixUp Training:
		
		
		Temperature Scaling: $\bar{Y}^{T} = \mathrm{Sharpen}(\hat{Y}^{T})$
		
		$X^{all}$ = Shuffle(Concat($\hat{X}^{T},X^{T}$))
		\FOR{$n = 1$ \TO $(N_{\hat{T}}*K+N_{T})/BATCHSIZE$ }
		\FOR{$b = 1$ \TO $BATCHSIZE$ } 
		\STATE{
			
			$r_{1} = Randint([1,BATCHSIZE])$, $r_{2} = Randint([1,BATCHSIZE])$
			
			$x_{b} = \lambda*x_{r_{1}}^{all} + (1 - \lambda) * x_{r_{2}}^{all}$
			
			$y_{b} = \lambda*y_{r_{1}}^{all} + (1 - \lambda) * y_{r_{2}}^{all}$
			
			$Append(X_{B},x_{b}), Append(Y_{B},y_{b})$
		} 
		
		\ENDFOR 
		
		Train classification network $h_{\hat{T}}(X^{B}, X^{T} | \theta_{\hat{T}})$ with $X^{T}$, $X^{B}$, $Y^{T}$, and $Y^{B}$. 
		\ENDFOR

	\end{algorithmic}
\end{algorithm}
\end{document}